\documentclass[10pt, conference, compsocconf]{IEEEtran}
\usepackage{graphicx}
\usepackage{mathptmx}
\usepackage{url}
\usepackage{xspace}
\usepackage{hyperref}

\def\hb{\hbox to 10.7 cm{}}

\makeatletter
\newcommand*\titleheader[1]{\gdef\@titleheader{#1}}
\AtBeginDocument{%
  \let\st@red@title\@title
  \def\@title{%
    \bgroup\normalfont\small\centering\@titleheader\par\egroup
    \vskip1.5em\st@red@title}
}
\makeatother

\newcommand{\euretile}{EURETILE\xspace}
\newcommand{\exanest}{ExaNeSt\xspace}
\newcommand{\euroexa}{EuroExa\xspace}
\newcommand{\hbp}{HBP\xspace}
\newcommand{\hbplong}{Human Brain Project\xspace}
\newcommand{\exanestgrant}{Grant Agreement No. 671553 (ExaNeSt)\xspace}
\newcommand{\hbpgrant}{Specific Grant Agreements No. 785907 (Human Brain Project SGA2) and No. 720270 (HBP SGA1)\xspace}
\newcommand{\euroexagrant}{Grant Agreement No. 754337 (EuroEXA)\xspace}

\newcommand{\nvidia}{NVIDIA\xspace}

\newcommand{\ie}{\textit{i.e.}\xspace}
\newcommand{\eg}{\textit{e.g.}\xspace}

\newcommand{\euroserver}{EuroServer\xspace}
\newcommand{\wavescales}{WaveScalES\xspace}
\newcommand{\dpsnn}{DPSNN\xspace}
\newcommand{\gbe}{GbE\xspace}
\newcommand{\lif}{Leaky Integrate-and-Fire\xspace}
\newcommand{\dpsnnlong}{Distributed and Plastic Spiking Neural Network\xspace}

\title{Real-time cortical simulations: energy and interconnect scaling on distributed systems}
\titleheader{Final peer reviewed version as: F. Simula et al., "Real-time cortical simulations: energy and interconnect scaling on distributed systems", 2019 27th Euromicro International Conference on Parallel, Distributed and Network-based Processing (PDP), Pavia, Italy, February 13-15, 2019, pp. 283-290. doi: 10.1109/EMPDP.2019.8671627}

\author{
  \IEEEauthorblockN{Francesco Simula}
  \IEEEauthorblockA{INFN Sezione di Roma\\ Rome,
    Italy\\ francesco.simula@roma1.infn.it}\\

  \IEEEauthorblockN{Elena Pastorelli}

  \IEEEauthorblockA{INFN Sezione di Roma and PhD Program in
    Behavioural Neuroscience, ``Sapienza'' University of Rome\\ Rome,
    Italy\\ elena.pastorelli@roma1.infn.it}\\

  \IEEEauthorblockN{Pier Stanislao Paolucci, Michele Martinelli, Alessandro Lonardo, Andrea Biagioni, Cristiano Capone, \\Fabrizio Capuani, Paolo Cretaro, Giulia De Bonis,
    Francesca Lo Cicero, Luca Pontisso, Piero Vicini}

  \IEEEauthorblockA{INFN Sezione di Roma\\ Rome,
    Italy\\ \{first-name.family-name\}@roma1.infn.it}\\

  \IEEEauthorblockN{Roberto Ammendola}

  \IEEEauthorblockA{INFN Sezione di Tor Vergata and Electronic
    Engineering Dept., University of Roma ``Tor Vergata''\\ Rome,
    Italy\\ roberto.ammendola@roma2.infn.it}

}
\author{\IEEEauthorblockN{Francesco Simula\IEEEauthorrefmark{1},
Elena Pastorelli\IEEEauthorrefmark{2},
Pier Stanislao Paolucci\IEEEauthorrefmark{1}, 
Michele Martinelli\IEEEauthorrefmark{1},
Alessandro Lonardo\IEEEauthorrefmark{1},\\
Andrea Biagioni\IEEEauthorrefmark{1},
Cristiano Capone\IEEEauthorrefmark{1},
Fabrizio Capuani\IEEEauthorrefmark{1},
Paolo Cretaro\IEEEauthorrefmark{1},
Giulia De Bonis\IEEEauthorrefmark{1},\\
Francesca Lo Cicero\IEEEauthorrefmark{1},
Luca Pontisso\IEEEauthorrefmark{1},
Piero Vicini\IEEEauthorrefmark{1} and
Roberto Ammendola\IEEEauthorrefmark{3}}
\IEEEauthorblockA{\IEEEauthorrefmark{1}INFN Sezione di Roma\\ Rome,
    Italy\\ \{first-name.family-name\}@roma1.infn.it}
\IEEEauthorblockA{\IEEEauthorrefmark{2}INFN Sezione di Roma and PhD Program in
    Behavioural Neuroscience, ``Sapienza'' University of Rome\\ Rome,
    Italy\\ elena.pastorelli@roma1.infn.it}
\IEEEauthorblockA{\IEEEauthorrefmark{3}INFN Sezione di Tor Vergata and Electronic Engineering Dept., University of Roma ``Tor Vergata''\\ Rome,
    Italy\\ roberto.ammendola@roma2.infn.it}}

\begin{document}

\pagestyle{headings}
\def\thepage{}

\maketitle

\begin{abstract}
We profile the impact of computation and \mbox{inter-processor}
communication on the energy consumption and on the scaling of cortical
simulations approaching the \mbox{real-time} regime on distributed
computing platforms.
Also, the speed and energy consumption of processor architectures
typical of standard HPC and embedded platforms are compared.
We demonstrate the importance of the design of \mbox{low-latency}
interconnect for speed and energy consumption.
The cost of cortical simulations is quantified using the Joule per
synaptic event metric on both architectures.
Reaching efficient \mbox{real-time} on large scale cortical
simulations is of increasing relevance for both future
\mbox{bio-inspired} artificial intelligence applications and for
understanding the cognitive functions of the brain, a scientific quest
that will require to embed large scale simulations into highly complex
virtual or real worlds.
This work stands at the crossroads between the \wavescales
experiment in the \hbplong (\hbp), which includes the
objective of large scale \mbox{thalamo-cortical} simulations of brain
states and their transitions, and the \exanest and \euroexa projects, that
investigate the design of an \mbox{ARM-based}, \mbox{low-power} High
Performance Computing (HPC) architecture with a dedicated interconnect
scalable to million of cores; simulation of deep sleep Slow Wave
Activity (SWA) and Asynchronous aWake (AW) regimes expressed by thalamo-cortical models are among their benchmarks.
\end{abstract}




\section{Introduction}

In modern HPC and embedded systems, the most constraining limits to scaling are
those related to power draw and dissipation.
In HPC the electricity bill is the main contributor to the total 
cost of running an application so that \mbox{energy-efficiency} 
is becoming a fundamental requirement for large scale platforms, 
and power is a critical design figure for any embedded system.
In this context, the feasibility of a computing system must not only pass through performance
assessment of processors but also their \mbox{performance-per-watt}
ratio.
Several scientific communities are exploring \mbox{non-traditional}
\mbox{many-core} processors architectures looking for a better
tradeoff between \mbox{time-to-solution} and
\mbox{energy-to-solution}.
Some architectures of this kind are the Graphics Processing Unit (GPU)
or those like the MPSoC that come from the embedded world, where
\mbox{ARM-based} \mbox{System-on-Chip} designs dominate the market of
\mbox{low-power} and \mbox{battery-powered} devices such as tablets
and smartphones.

A number of research projects are active in trying to design an actual
HPC platform along this direction.
The \mbox{Mont-Blanc} project~\cite{montblanc:2016:short,montblanc:2017:Online},
coordinated by the Barcelona Supercomputing Center, has deployed two generations
of HPC clusters based on ARM processors, developing also the corresponding
ecosystem of HPC tools targeted to this architecture.
Another example is the \mbox{EU-FP7}
\euroserver~\cite{Marazakis:EUROSERVER:2016} project, coordinated by
CEA, which aims to design and prototype technology, architecture, and
systems software for the next generation of datacenter
``microservers'', exploiting \mbox{64-bit} ARM cores.

Unraveling how the brain works is a formidable scientific and HPC
undertaking.
The human brain includes about $10^{15}$ synapses and $10^{11}$
neurons activated at a mean rate of several Hz; as a digital
simulation, it is a significant coding challenge and has very exacting
requirements for an adequate computing architecture, even at the
highest abstraction level.

Fast simulation of spiking neural network models plays a dual role:
(i) it contributes to the solution of a scientific grand challenge ---
\ie the comprehension of brain activity --- and, (ii) by including it
into embedded systems, it can enhance applications such as autonomous
navigation, surveillance and robotics, requiring \mbox{real-time}
performances.
Moreover, \mbox{real-time} simulation of neural networks will be
essential for understanding the mechanisms underlying the cognitive functions of the brain.
Indeed, brain simulations should be embedded in complex environments,
\eg robotic platforms interacting with the world in \mbox{real-time},
which makes requirements on power consumption so much tighter.
Therefore, cortical simulations assume a driving role in shaping the
architecture of either specialized and \mbox{general-purpose}
\mbox{multi-core}/\mbox{many-core} systems to come, standing at the
crossroads between embedded and HPC.
See, for example~\cite{Merolla668:short}, describing the TrueNorth
\mbox{low-power} specialized hardware architecture dedicated to
embedded applications, and~\cite{Stromatias:2013} discussing the power
consumption of the SpiNNaker hardware architecture, based on embedded
\mbox{multi-cores}, dedicated to brain simulation.
Worthy of mention are also~\cite{gewaltig:2007,Modha:2011} as examples
of approaches based on standard HPC platforms and
\mbox{general-purpose} simulators.

The \wavescales experiment in the \hbplong (\hbp)
has the goal of matching experimental measures with simulations of
Slow Wave Activity (SWA) during deep sleep and anaesthesia, the
transition to other brain states, and the interplay between cortical waves and memories with a focus in developing
dedicated, parallel/distributed technologies able to overcome some of
the limits faced by current attempts at brain simulation.
On a different line of research, the
\exanest~\cite{DSD:EXANEST:2016:short} and \euroexa projects
investigates the design of an \mbox{ARM-based}, \mbox{low-power} High Performance Computing (HPC) architectures with dedicated interconnects
scalable to million of cores; thalamo-cortical 
simulations are among their motivating benchmarks.
At the joint between these projects stands the \dpsnnlong
(\dpsnn) simulation engine, developed by the APE
Parallel/Distributed Computing Laboratory at INFN; its C/C++ code is
written according to the MPI \mbox{multi-process} paradigm and is
designed to be easily portable to exotic architectures and to stress
either the available networking or computing resources.
In this paper, the simulator is used to compare the scaling in time
and energy consumption of an \mbox{Intel-based} HPC cluster ---
equipped with \mbox{high-performance} InfiniBand connectivity in
addition to ordinary Ethernet --- with that of different
\mbox{ARM-based} platforms, taken as representatives of the class of
new, \mbox{low-power} HPC systems like those pursued by
\exanest and \euroexa.

\begin{figure}[!b]
  \includegraphics[width=.48\textwidth]{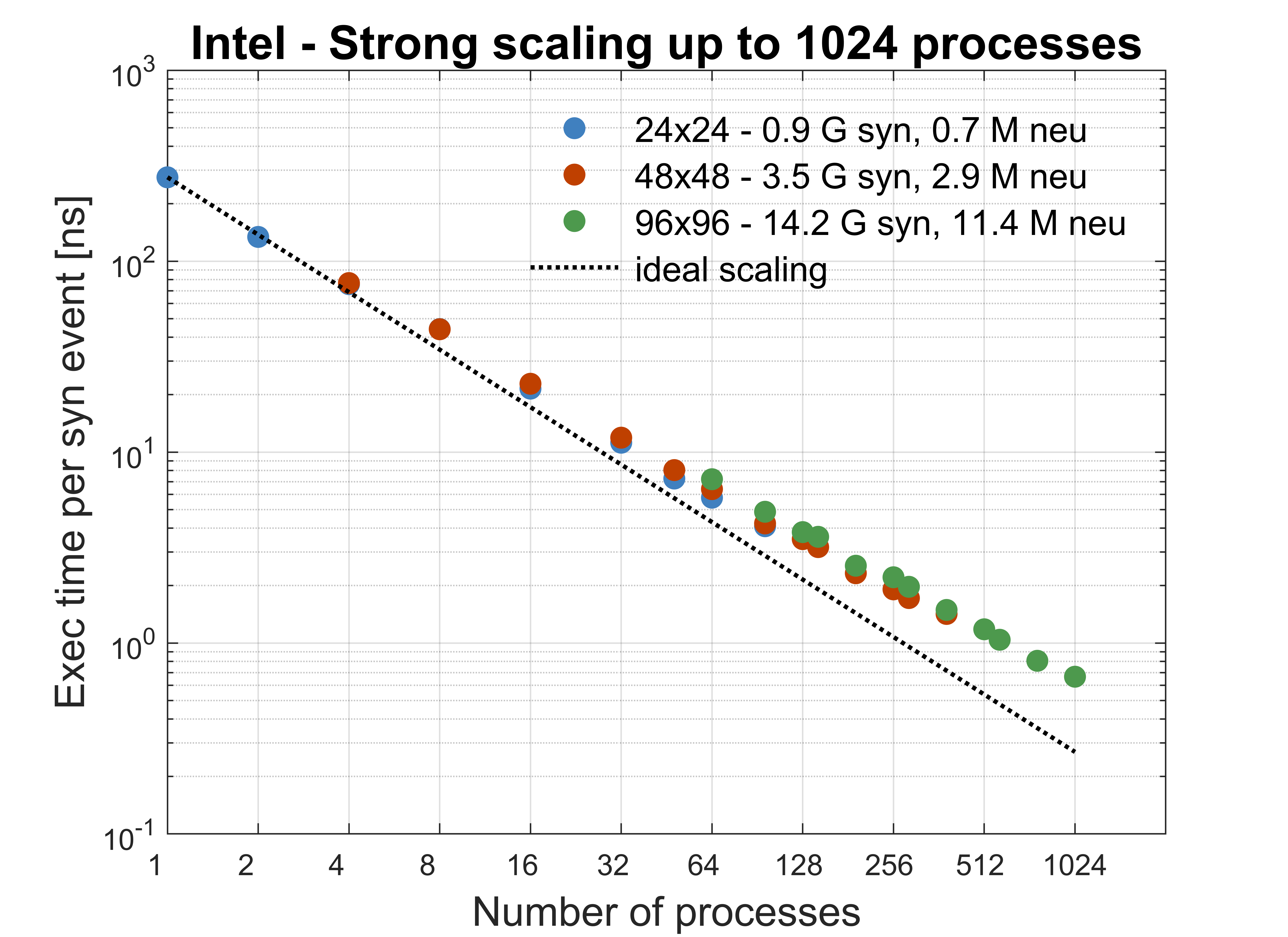}
  \caption{Strong scaling up to 1024 processes of large neural
    networks on an \mbox{IB-equipped} \mbox{Intel-based} cluster.}
  \label{fig:IntelScaling1024}
\end{figure}

In previous works, we demonstrated the scalability of our simulation engine up to 1K
processes~\cite{pastorelli:PDP2018:gaussianLateral} when
applied to realistic long range synaptic connectivity~\cite{schnepel:2015} 
and described its internal architecture~\cite{Paolucci:2013:Distributed}.
Within \hbp the simulator has been applied to the study of Slow
Waves
Activity~\cite{caponeSanchezvivesDelGiudiceMattia:2017,ruiz:2011,stroh:2013}
in large scale cortical fields (up to 14 billion synapses), a
different HPC challenge in which \mbox{real-time} is not required
(Figure~\ref{fig:IntelScaling1024}).

Originally, this simulation engine was developed as a \mbox{mini-application}
benchmark in the framework of the \euretile FP7
project~\cite{EURETILE:JSA:2016:short}.

The code organization and its compactness endow our application with a
high degree of tunability and with it the chance of testing different
areas of the executing platforms; by varying the number of neurons per
core in the simulated neural net, the analysis can be moved from the
performances of the platform interconnect --- with relatively few neurons, each one projecting thousands
of synapses --- to the computing and memory resources --- with more
neurons per core.
Full biological realism of a cortical tissue would require a number of synapses per neuron in the 
range between 5000 and 10000. 
Plus, the representation of large scale cortical systems needs the projection 
of long range \mbox{intra-areal} sparse connectivity described either by 
distance and layer dependent probability rules or by explicit lists of connections. 
\mbox{Inter-areal} connectivity is instead derived from the description of the 
sparse \mbox{long-range} connectome. 
Both kind of synaptic adjacency matrices depend on the spatial location 
of source and target neurons. 
If a large scale neural network is distributed on a grid of processes 
using a spatial mapping (i.e. a set of neighbouring neurons and incoming synapses 
is assigned to each process), the transport of spiking messages carried by the sparse 
synaptic adjacency matrix does not typically require an \mbox{all-to-all} 
interconnection between processes. 
Indeed, we demonstrated the advantages of such a reduction of the adjacency matrix 
between processes for the scaling of simulations of large networks with biologically plausible 
\mbox{intra-areal} \mbox{long-range} connections in~\cite{pastorelli:PDP2018:gaussianLateral}.
However, the execution time for such large scale systems (Figure~\ref{fig:IntelScaling1024}) 
is still one or two order of magnitude slower
than the real time domain we focus on in this paper. 
If smaller number of neurons are considered, as necessary to reduce the 
network size to compatibility with real time execution, 
the sparsity of the synaptic adjacency matrix would also be reduced 
and the simulation would typically require \mbox{all-to-all} 
interprocess communications.
As we will see, the communication of spikes among neurons 
is dominated by latency for network sizes in the explored range 
on contemporary HPC and embedded platforms. 
Therefore, we adopted in this paper a simple synaptic adjacency matrix:
an homogeneous connection probability that simplifies the analysis of the scaling behaviour.
Finally, we reduced the number of synapses per neuron to 1125, this way further stressing intercommunication latency (moderate size of payloads) and enabling the simulation of networks with a few more neurons, 
with a potentially higher representational power, but still needing the support of \mbox{all-to-all} \mbox{inter-process communication}.

This paper addresses the measure of power consumption and \mbox{energy-to-solution} for \mbox{real-time} cortical simulation and profiles the relative scaling of computation, communication and
synchronization.
Specifically, we perform a number of neural
simulations to compare the performances of ARM- and \mbox{Intel-based}
\mbox{multi-core} platforms, with further focus on the possible impact of the
usage of \mbox{off-the-shelf} vs. custom networking components.

\section{\mbox{Mini-application} benchmarking tool}
Evaluation of HPC hardware is a key element especially in the first
stages of a project --- \ie definition of specification and design ---
and during the development and implementation.
Key components impacting performance should be identified in the early
stages of the development, but full applications are too complex to be
run on simulators and hardware prototypes.
In usual practice, hardware is tested with very simple kernels and
benchmarking tools which often reveal their inadequacy as soon as they
are compared with real applications running on the final platform,
showing a huge performance gap.

In the last years, a new category of compact, \mbox{self-contained}
proxies for real applications called \textit{mini-apps} have appeared.
Although a full application is usually composed by a huge amount of
code, the overall behaviour is driven by a relatively small subset of it.
\mbox{Mini-apps} are composed by these core operations providing a
tool to study different subjects:
(i) analysis of the computing device --- \ie the node of the system.
(ii) evaluation of scaling capabilities, configuring the
\mbox{mini-apps} to run on different numbers of nodes, and
(iii) study of the memory usage and the effective throughput towards
the memory.

This effort is led by the Mantevo project~\cite{heroux:2009}, that
provides application performance proxies since 2009.
Furthermore, the main research computing centers provide sets of
\mbox{mini-applications}, adopted when procuring the systems, as in
the case of
the \mbox{NERSC-8}/Trinity Benchmarks~\cite{Cordery:2014}, used to
assess the performance of the Cray XC30 architecture, or
the \textit{Fiber Miniapp Suite}~\cite{fiber}, developed by RIKEN Advanced
Institute for Computational Science (RIKEN AICS) and the Tokyo
Institute of Technology.

In this work, we used \dpsnn as a \mbox{mini-application}
benchmarking tool to simulate networks of \mbox{point-like} spiking
neurons of size compatible with reaching the real-time target.
The network is composed of 80\% \lif neurons with Spike Frequency Adaptation (SFA), 
representing cortical pyramidal excitatory neurons with fatigue and 20\% inhibitory neurons. 
SFA is switched off for inhibitory neurons.  
This network is a down-scaling of a grid of cortical columns~\cite{pastorelli:PDP2018:gaussianLateral}
with realistic long range inter-columnar synaptic connectivity~\cite{schnepel:2015}. 
This network is able to enter both an asynchronous awake-like regime and a 
deep-sleep-like slow wave activity, by tuning the values of SFA and stimulation.
Within the Wavescales experiment, a similar model with SFA 
is extended to study the interactions between Slow Waves Activity, 
memory association and synaptic homeostasis in a thalamo-cortical model
applied to the classification of MNIST handwritten digits~\cite{Capone:2018:SleepMemAssSynHom}.
In this paper, synapses inject instantaneous \mbox{post-synaptic} currents while
synaptic plasticity is disabled.
The simulator implements a mixed \mbox{event-driven} (synaptic and
neural dynamics) and \mbox{time-driven} (exchange of spiking messages)
integration scheme.
As discussed in the previous section, the number of synapses projected by
each neuron is kept constant with an average value of 1125 synapses
per neuron, the synaptic adjacency matrix is homogenously sparse, 
and neurons are evenly distributed among processes.

Each neuron receives also the stimulus of 400 ``external'' synapses,
each one delivering a Poissonian spike train at a rate of about 3~Hz.
After an initial transient, the neural network enters an asynchronous
irregular firing regime at a mean rate of about 3.2~Hz in all simulations 
used for the scaling measures of this paper.

\mbox{Inter-process} communication is necessary to deliver spikes to
target neurons residing on a process different from the one hosting
the source neuron.
Spikes are delivered using the AER representation (spiking neuron ID,
emission time)~\cite{lazzaro:1993}; in our case 12 byte per spike are
required.
The exchange of spikes is implemented in the set-up of this paper by means of
synchronous MPI collectives.
In a process, all spikes produced by neurons and targeted to neurons
belonging to another are packed into a single message and delivered.
The total number of messages required for \mbox{all-to-all} communication
increases with the square of the number of processes on which the simulation is run.
This throttles the application into different regimes, allowing to
stress and test several elements of the execution platform.

Here is a rundown of the application tasks that the simulator performs
and that allow to gauge the components of the architecture under test:
\begin{itemize}
\item \textbf{Computation}: \mbox{event-driven} integration of all
  neural dynamics and synaptic current injection events, occurring in
  a single network synchronization time step (set to 1~ms). This
  includes a component dominated by memory access to: 1- time delay
  queues of axonal spikes, 2- lists of \mbox{neuro-synaptic}
  connections, 3- lists of synapses.
\item \textbf{Communication}: transmission along the interconnect
  system of the axonal spikes to the subset of processes where target
  neurons exist (in the specific set-up of this paper, all processes).
\item \textbf{Synchronization}: synchronization barrier inserted to
  simplify the weighting of computation and communication components.
\end{itemize}

Fluctuations in computation load or communication congestion cause
idling cores and diminished parallelization.
The relative weight of computation increases with the number of
incoming synapses per process.
On the other side, a higher number of processes results in higher
relative communication costs.

\section{Scaling towards \mbox{real-time}}
\label{sec:scaling_real_time}

In this domain, being ``real-time'', under a ``soft'' assumption, means a work point for the
application such that the total \mbox{wall-clock} time for running it is not
greater than the total simulated time, a condition necessary, but not sufficient, for robotics
applications and embedding HPC simulations into virtual or real world
environments, that would impose more stringent ``hard'' constraint to be satisfied at the scale of each step, lasting at most a few tens of ms each. 
The aim of this work is to identify the obstacles that impede reaching
the \mbox{real-time} target for large neural networks.
We performed a set of strong scaling tests on neural networks of increasing size executed on
both Intel and \mbox{ARM-based} distributed platforms. 
For all network, we simulated 10~s of neural activity.
The first testbed we used, representative of standard HPC systems,
is made of Intel Xeon E5-2630 v2 processors (clocked at 2.60GHz) 
communicating over a \mbox{ConnectX-class} InfiniBand interconnect. 
For what concern the case of ARM-based distributed systems, representative of the
embedded system world, we used two different testbeds, respectively based
on Trenz and Jetson boards, detailed later in this section.

\begin{figure}[!hbt]
  \includegraphics[width=.48\textwidth]{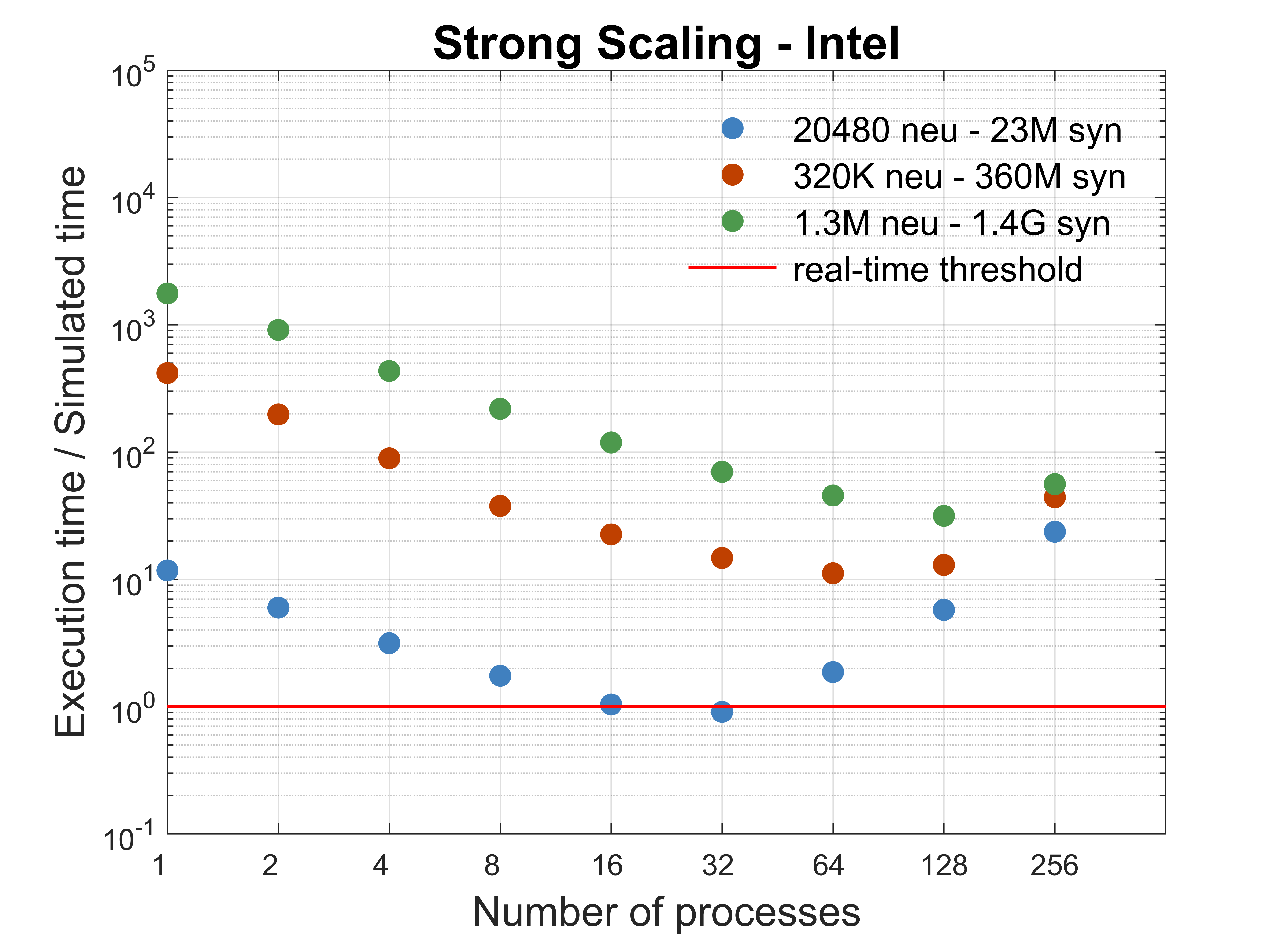}
  \caption{Strong scaling of different problem sizes on an
    \mbox{IB-equipped} \mbox{Intel-based} platform. The red line is the
threshold to be reached for soft \mbox{real-time} execution.}
  \label{fig:IntelScaling}
\end{figure}

\begin{figure}[!hbt]
  \includegraphics[width=.48\textwidth]{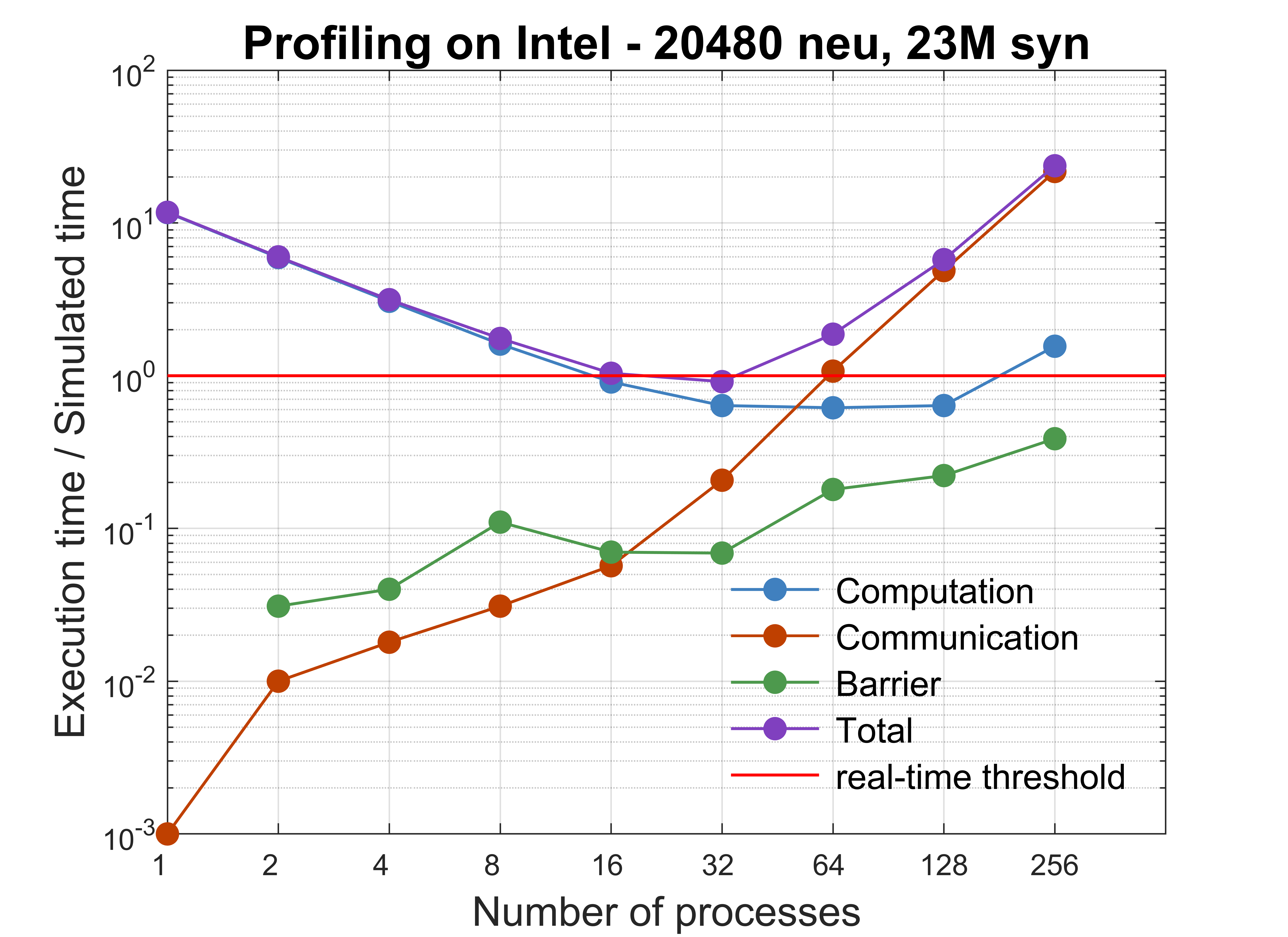}
  \caption{\dpsnn analysis of the \mbox{Intel-based} platform.}
  \label{fig:IntelProfile}
\end{figure}

Figure~\ref{fig:IntelScaling} shows the runtimes for three neural network sizes.
They should all be able to run in \mbox{real-time} if
the scaling valid for larger configurations applied (see Figure~\ref{fig:IntelScaling1024}).
Indeed, the 20480 neurons configuration reaches \mbox{real-time}
(9.15 seconds to simulates 10 seconds of activity). 
The network with 20480 neurons reached its maximum speed when distributed on $32$ processes (Figure~\ref{fig:IntelScaling}).
Communication and synchronization are the main obstacles against scaling (see Figure~\ref{fig:IntelProfile} and Table~\ref{tab:dpsnntask}).
For the 20480 neuron configuration they 
block a further acceleration over $32$ processes and start impeding the scaling toward \mbox{real-time} of 
larger neural networks after a threshold corresponding to a larger number of processes for the configurations with 320K neurons ($16 \times$ the 20480 network) and 1280K neurons ($64 \times$).

\begin{table}[!hbt]
\footnotesize
\centering
\tabcolsep=0.11cm
\caption{Profiling of execution components for different network sizes.}
\label{tab:dpsnntask}
\begin{tabular}{ |l| *{7}{c|} }
  \hline
  Neurons     & \multicolumn{3}{c|}{20480N}   & \multicolumn{2}{c|}{320KN}    & \multicolumn{2}{c|}{1280KN}   \\ \hline
  Synapses    & \multicolumn{3}{c|}{2.30E+07} & \multicolumn{2}{c|}{3.60E+08} & \multicolumn{2}{c|}{1.44E+09} \\ \hline
  Procs       &   4    & 32           & 256             &   4             & 256              &    4            & 256              \\ \hline
  Wall-clock (s)  &  31.5 & \textbf{9.15}             & 237          & 893         & 441           & 4341        & 561           \\ \hline
  Computation     & \textbf{97.6}\%  & 69.7\%   &   6.6\%         & \textbf{98.1}\% &  21.7\%          & \textbf{99.4}\% &  50.0\%          \\ \hline
  Communicat.    &   0.6\%   & 22.7\%           & \textbf{91.7}\% &   0.1\%         & \textbf{79.9}\%  &    0.1\%        & \textbf{48.1}\%  \\ \hline
  Barrier         &   1.3\%   & 7.5\%         &   1.6\%         &   1.8\%         &   1.1\%          &    0.5\%        &   1.9\%          \\ \hline
\end{tabular}
\end{table}

In our simulation, the network communicates spikes every simulated
millisecond, the payload for each spike is 12 byte and the average
firing rate is about 3 Hz.
As a consequence, when the number of cores increases, the network
produces a very large number of small message packets.
Therefore, this test highlights a ``latency'' limitation of the
interconnect.
In general, commercial \mbox{off-the-shelf} interconnects offer
adequate throughput when moving large amounts of data but typically
trudge when the communication is \mbox{latency-dominated}.
This issue with communication --- manifesting here with a number of
computing cores which is, by today's standards, not large --- is
similar to that encountered by the parallel cortical simulator
C2~\cite{Ananthanarayanan:2009} --- targeting a scale in excess of
that of the cat cortex --- on the Dawn Blue Gene/P supercomputer at
LLNL, with 147456 CPUs and 144~TB of main memory.
The capability to replicate the behaviour of a supercomputer with a
\mbox{mini-app} running on a limited number of 1U servers hints at an
interesting performance improvement at both larger scale and smaller
\mbox{real-time} configuration if identified obstacles to scaling were
removed.

Similar results are obtained performing the same test on two
\mbox{ARM-based} platforms; one is the \mbox{ARM-based} prototype of
the \exanest project~\cite{KATEVENIS:MICRO:2018} and the other is a
commercial development board by \nvidia equipped with an ARM SoC
(Jetson TX1).

\begin{figure}[!hbt]
  \includegraphics[width=.48\textwidth]{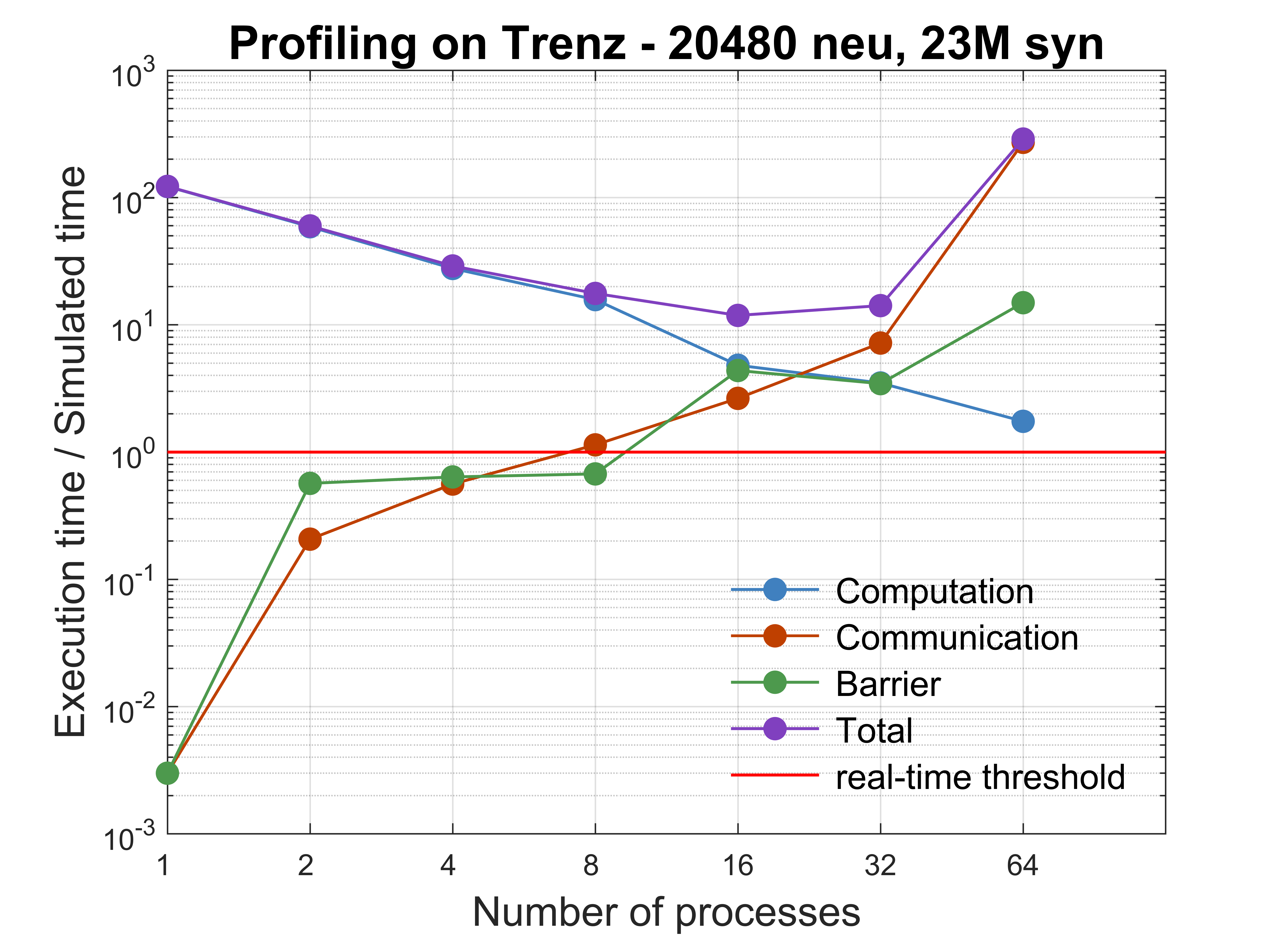}
  \caption{\dpsnn analysis of the Trenz platform.}
  \label{fig:TrenzProfile}
\end{figure}

\begin{figure}[!hbt]
  \includegraphics[width=.48\textwidth]{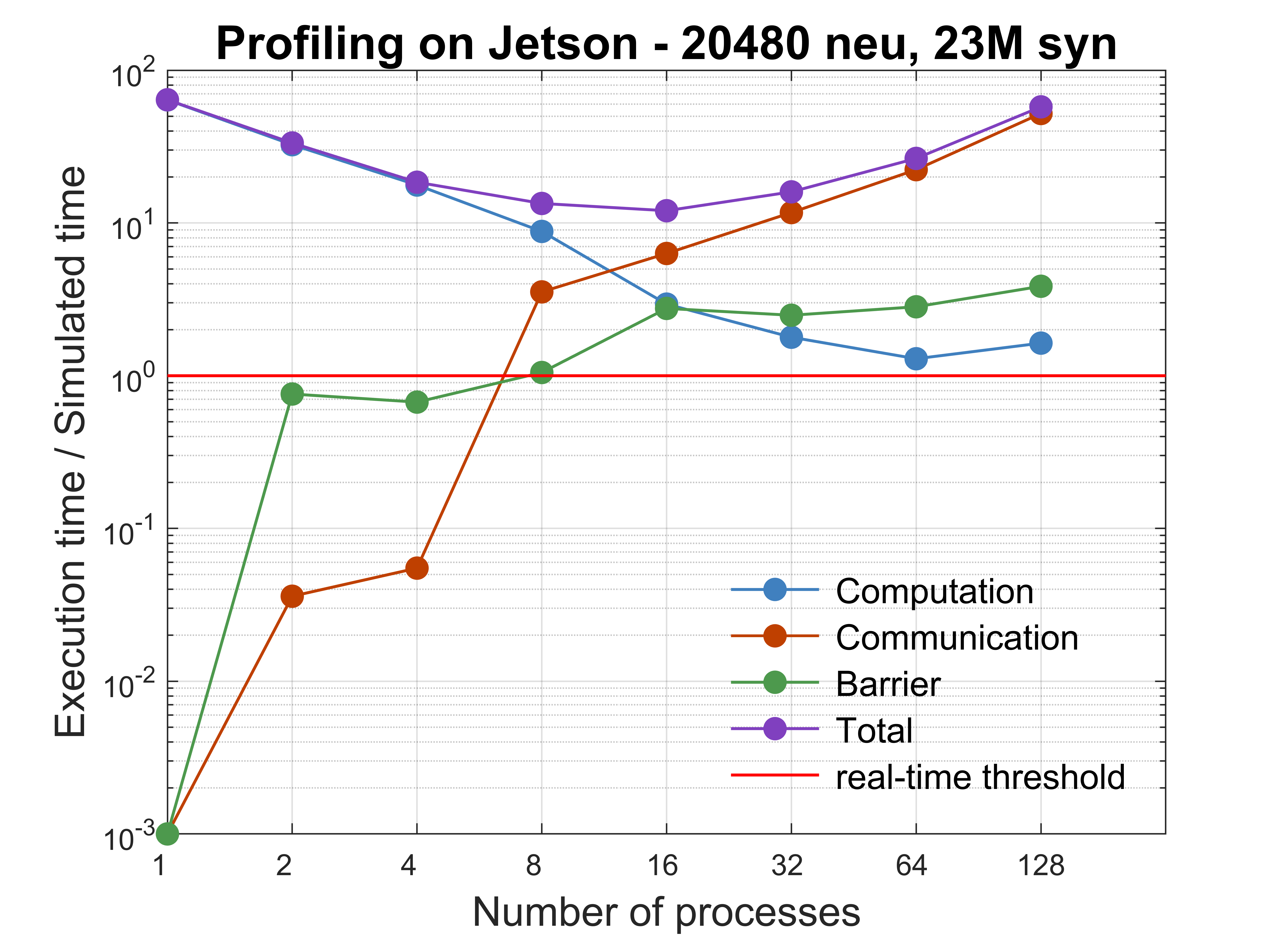}
  \caption{\dpsnn analysis of the \nvidia SoC platform.}
  \label{fig:JetsonProfile}
\end{figure}

The \exanest prototype is composed by four nodes, each node consisting
of a TEBF0808 Trenz board equipped with a Trenz TE0808 UltraSOM+
module.
The Trenz UltraSOM+ consists of a Xilinx Zynq UltraScale+
\mbox{xczu9eg-ffvc900-1-e-es1} MPSoC and 2~Gbytes of DDR4 memory.
The Zynq UltraScale+ MPSoC incorporates both a processing system
composed by \mbox{quad-core} ARM \mbox{Cortex-A53} and the
programmable logic --- left unused in this test.
All four nodes are connected together through a 1~Gbps
\mbox{Ethernet-based} network.
Given that the available cores are limited to 16, the scaling was
pushed further up by using the ``heterogeneous'' mode of MPI, which
allows launching an application as a single MPI instance that
simultaneously uses distinct executables for different architectures;
in this way, the simulation of the neural network is split between partitions
of processes executing on ARM and Intel cores.
A similar partitioning approach has been used also for the platform based on
Jetson boards.
Intel cores are about ten times faster than the ARMs on the Trenz boards
and about 5 times faster than those on the Jetson (see execution times
for $1$, $2$ and $4$ processes in Figures~\ref{fig:IntelProfile}, 
\ref{fig:TrenzProfile} and \ref{fig:JetsonProfile} in a scaling regime
dominated by computation).
Therefore, the Intel ``bath'' of processes executed on the Intel partition
does not slow down the execution of the ARM Trenz and Jetson boards embedded in it.

\begin{figure}[!hbt]
  \includegraphics[width=.48\textwidth]{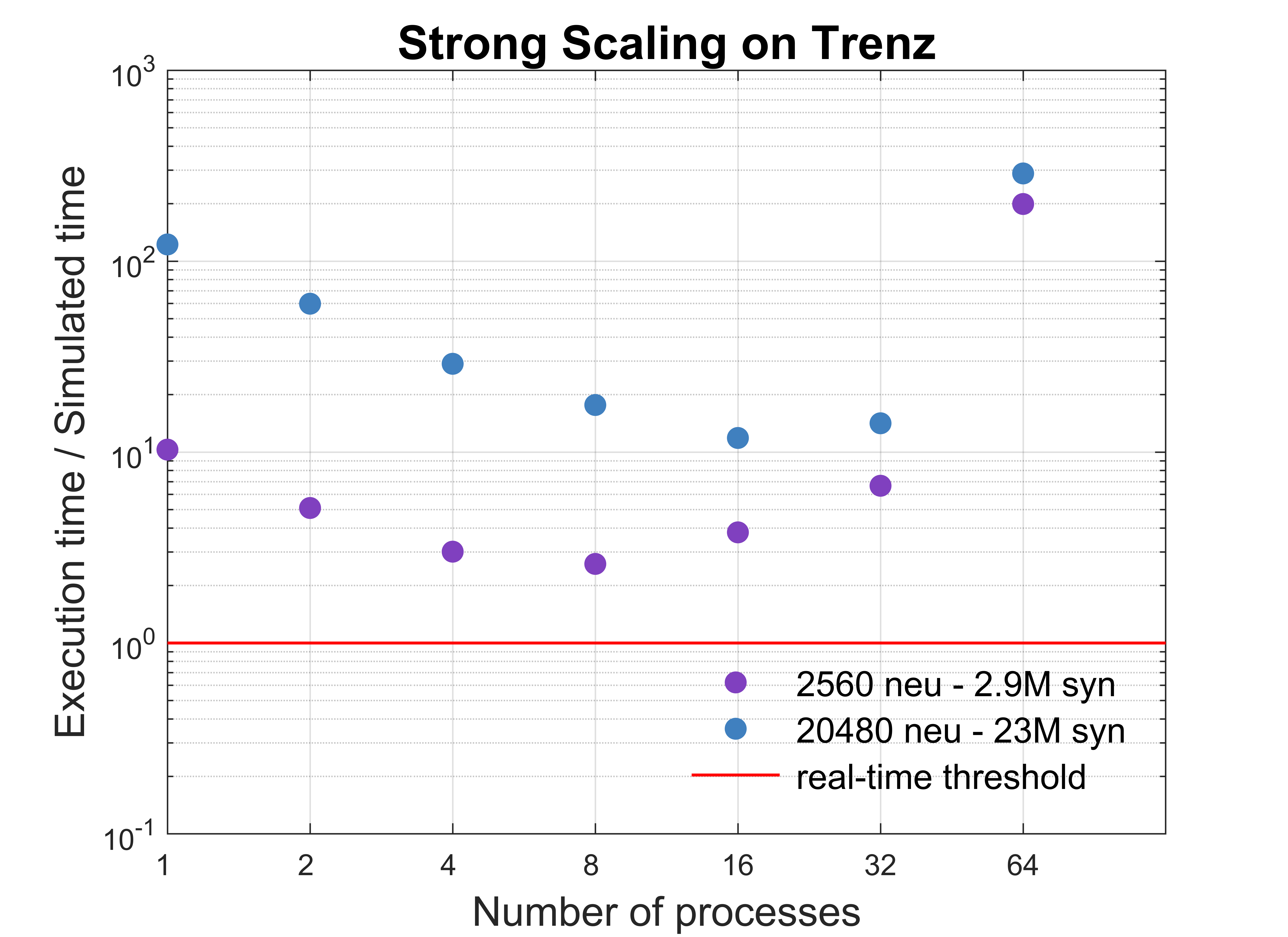}
  \caption{Strong scaling of a grid simulated on the Trenz platform
    equipped with \gbe interconnect.}
  \label{fig:TrenzScaling}
\end{figure}

The scaling of the system including the Trenz boards up to 64 processes
and the profiling of the computation, communication and synchronization 
components are reported in Figures~\ref{fig:TrenzScaling} and ~\ref{fig:TrenzProfile}.

The very same test was performed on two \nvidia Jetson TX1 boards
connected by an Ethernet 1~Gbit/s switch to emulate a
\mbox{dual-socket} node, each equipped with four ARM
\mbox{Cortex-A57}@2~GHz cores plus four ARM \mbox{Cortex-A53}@1.5~GHz
cores in 20~nm CMOS technology in big.LITTLE configuration; results
are in Figure~\ref{fig:JetsonProfile}.

\section{\mbox{Energy-to-Solution} analysis}

We estimate and compare the instantaneous power, total energy
consumption, execution time and energetic cost per synaptic event of a
spiking neural network simulator distributed on MPI processes running
the \dpsnn simulation engine on both the \mbox{low-power} and
standard HPC platforms.

The measures were performed with AC/DC current readings by a
\mbox{high-precision} GW Instek \mbox{GDM-8351} digital multimeter
connected via USB to a PC: for the SoCs the DC current was sampled
downstream the power supply --- rated as 19V DC output --- as long as
only one board was used; for two SoC boards and for the Intel servers
the AC current was sampled between the power strip feeding the
systems' plugs and the mains outlet --- rated as 220V AC output.
Such difference should not affect significantly the results, given the
closeness to one of the $\cos \varphi$ factor of the server power
supply.


The traditional computing system --- \ie ``server platform'' --- is
based on SuperMicro \mbox{X8DTG-D} 1U \mbox{dual-socket} servers
equipped with a mix of Xeon computing cores in the 32~nm CMOS
technology of the \mbox{Westmere-family}, \ie \mbox{exa-core}
X5660@2.8~GHz and \mbox{quad-core} E5620@2.4~GHz.
This ``server platform'' is juxtaposed to a typical ``embedded
platform'', which is composed by the two Jetson boards.

The ``embedded platform'' has 4~GB --- 1~GB per core --- of LPDDR4
memory, with a memory bandwidth declared as 25.6~GB/s; the ``server
platform'' has a varying amount of DDR3 memory (operating at 1333~MHz)
per node --- amounting to 1.5$\div$4~GB per core --- and a max
declared bandwidth of 32~GB/s.

\begin{figure}
  \centering \includegraphics[width=.48\textwidth]{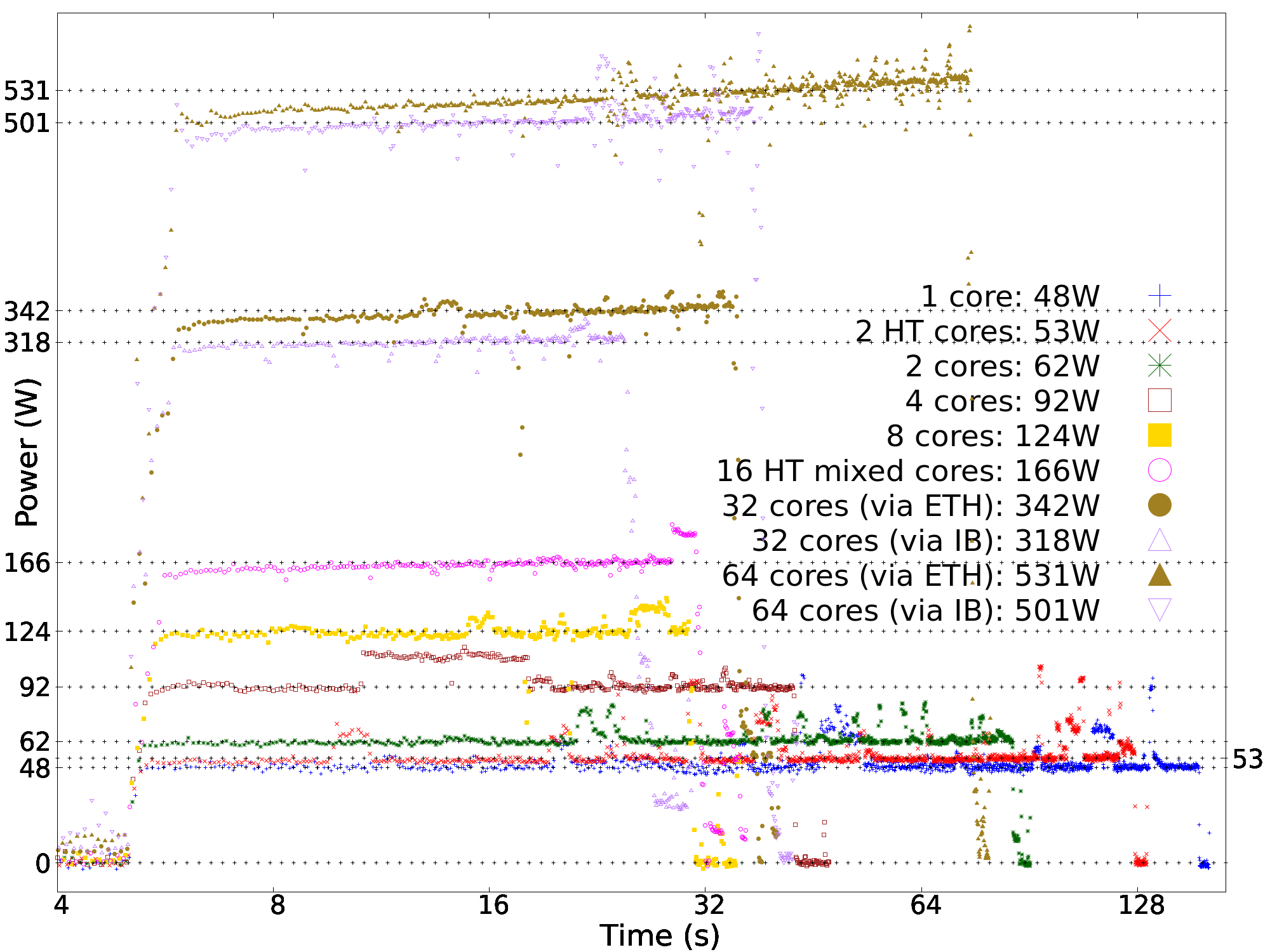}
  \caption{Scaling of the total power consumption on x86: the same 
total simulation workload is executed on a number of cores
doubling from 1 to 64. Temporal axis in logarithmic scale.}
  \label{fig:x86_power}
\end{figure}

\begin{figure}
  \centering
  \includegraphics[width=.48\textwidth]{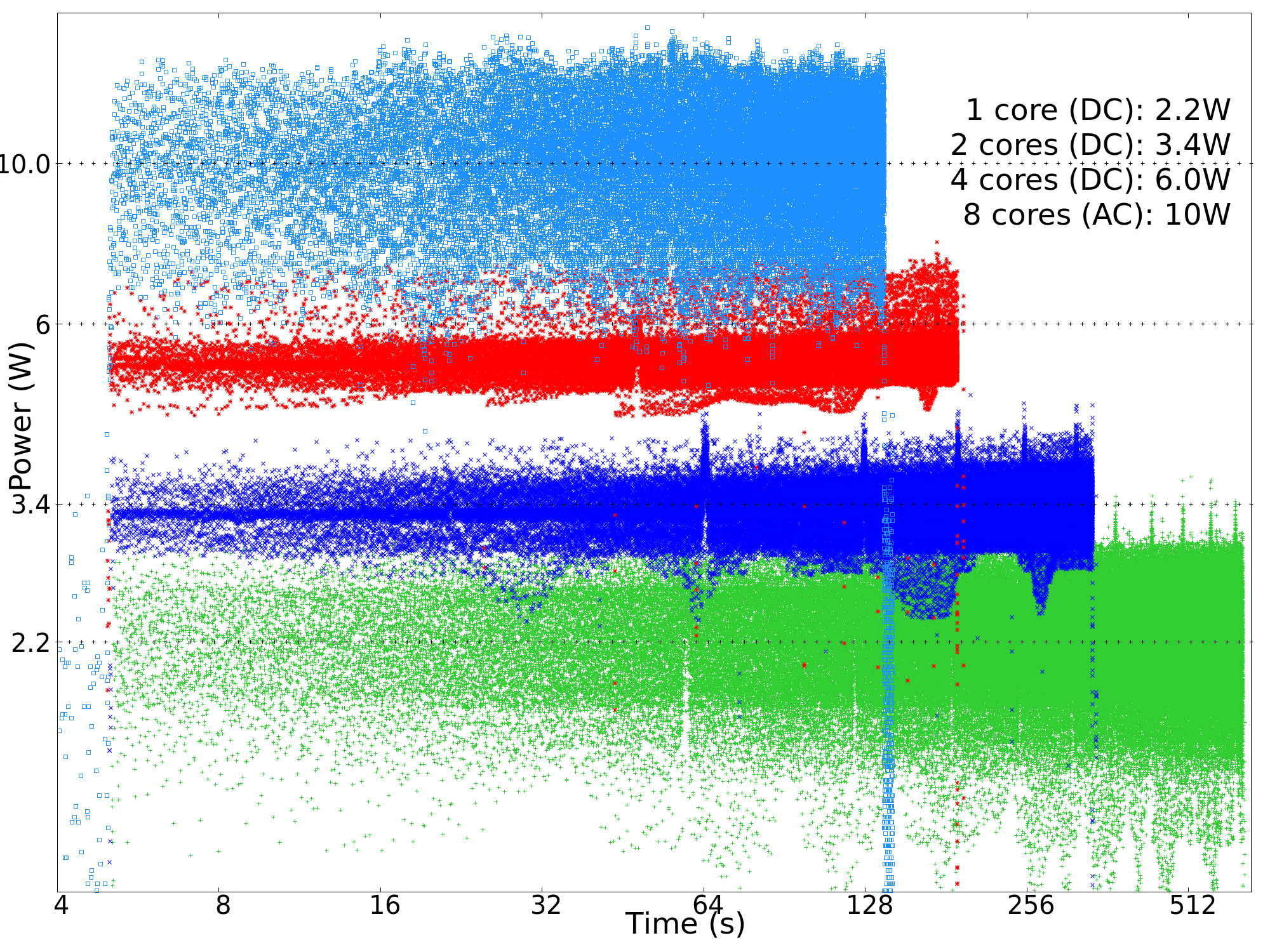}
  \caption{Power consumption on ARM. Temporal axis in logarithmic scale.}
  \label{fig:ARMpower}
\end{figure}

Power and energy consumption were obtained simulating 10~s of activity
of a network including $\sim$20480 neurons.
The results of a strong scaling test are reported in
Figure~\ref{fig:x86_power} for the ``server platform'' and in
Figure~\ref{fig:ARMpower} for the ``embedded'' one.
In both plots, the legend reports the number of processes employed.
Elapsed time is on the \mbox{X-axis} and the power draw is on the
\mbox{Y-axis}, the meter reading subtracted from a baseline that is
inferred by inspecting the plateau at application start, where 5~s of
artificial pause was inserted in the application.
Immediately after, a steep knee signals the real start of the
simulation and the final drop marks its end.
Note that, for the ``embedded platform'', the plot is split in two
ranges: measures between one and four cores are performed on a single board;
while two boards are used for eight cores.
For both measures we used only one multimeter; attaching the probe at the output 
of a single power supply - which is DC - was the approach used for a single board.
For two boards, we put the multimeter upstream the two power supplies, that implies 
an AC measure.
The transformers' draw causes a significantly higher baseline while
the readings are clearly noisier and more spread out.
These baselines stand at 564~W for the ``server platform'' --- from
Figure~\ref{fig:x86_power} --- and 49.2~W for the ``embedded
platform'' --- upper range of Figure~\ref{fig:ARMpower}.

\mbox{Energy-to-solution} and execution times for the Intel based
platform, computed using data from Figure~\ref{fig:x86_power}, are
summarized in Table~\ref{tab:x86_E2S}, while those from
Figure~\ref{fig:ARMpower} (ARM platform), are in
Table~\ref{tab:ARM_E2S}.

\begin{table}[!hbt]
\footnotesize
\centering
\caption{\dpsnn time, power and energy to solution on x86.}
\label{tab:x86_E2S}
\begin{tabular}{|c|c|c|c|}
\hline
\textbf{x86 cores}   & \textbf{Time (s)}    & \textbf{Power (W)} & \textbf{Energy to solution (J)} \\
\hline
\textbf{1}           & 150.9                & 48                 & 7243.2                          \\
\hline
\textbf{2 HT}        & 121.8                & 53                 & 6455.4                          \\
\hline
\textbf{2}           & 80.7                 & 62                 & 5003.4                          \\
\hline
\textbf{4}           & 37.4                 & 92                 & 3440.8                          \\
\hline
\textbf{8}           & 25.3                 & 124                & 3137.2                          \\
\hline
\textbf{16}          & 26.1                 & 166                & 4332.6                          \\
\hline
\textbf{32 plus ETH} & 30.0                 & 342                & 10260.0                         \\
\hline
\textbf{32 plus IB}  & 19.7                 & 318                & 6264.6                          \\
\hline
\textbf{64 plus ETH} & 69.3                 & 531                & 36798.3                         \\
\hline
\textbf{64 plus IB}  & 32.1                 & 501                & 16082.1                         \\
\hline
\end{tabular}
\end{table}

\begin{table}[!hbt]
\footnotesize
\centering
\caption{\dpsnn time, power and energy to solution on ARM.}
\label{tab:ARM_E2S}
\begin{tabular}{|c|c|c|c|}
\hline
\textbf{ARM cores}   & \textbf{Time (s)}    & \textbf{Power (W)} & \textbf{Energy to solution (J)} \\
\hline
\textbf{1}           & 636.8                & 2.2                & 1273.6                          \\
\hline
\textbf{2}           & 334.1                & 3.4                & 1135.9                          \\
\hline
\textbf{4}           & 185.0                & 6.0                & 1110.0                          \\
\hline
\textbf{8}           & 133.8                & 10                 & 1338.0                          \\
\hline
\end{tabular}
\end{table}

A peculiar corner case is relative to the second row of
Table~\ref{tab:x86_E2S}, where one physical core was used as two
HyperThreaded (HT) cores to host two MPI processes; the scaling is
clearly not as good as using two real, physical cores (third row), but
a small gain is attained nonetheless using what is fundamentally a
single core, which, at least for \dpsnn, does not completely
rule out using HyperThreading as often advised for general HPC
applications.
The system hosted on a single cluster node can use only up to 16
cores; as can be seen, the minimum \mbox{time-to-solution} is reached
with 32 cores, \ie 2 nodes, and only when choosing a
\mbox{low-latency} transport for the \mbox{inter-node} communication
such as InfiniBand --- ``IB'' in Table~\ref{tab:x86_E2S} and in
Figure~\ref{fig:x86_power} --- as opposed to Ethernet --- marked as
``ETH''.
Moreover, InfiniBand has another significant difference compared with
Ethernet; it draws measurably less power when in operation
($\sim$30~W), as the two branches of the \mbox{32-core} and the
\mbox{64-core} cases show.

Another interesting result is that the absolute minimum for
\mbox{energy-to-solution} at 8 cores requires not even using remote
communication, which is comprehensible given the relatively small size
of the problem being simulated.

\section{Conclusion}

The computational cost of neural simulations is approximately
proportional to the number of synaptic events.
The total number of synaptic events is the product of the number of
neurons, the number of synapses per neuron, the average firing rate
and the total simulation time.
The power efficiency can therefore be estimated with a J per synaptic
event metric by dividing the total \mbox{energy-to-solution} by the
total number of synaptic events.
As a reference, using this metric, the energetic cost of
Compass~\cite{Compass:short} --- an optimized simulator for the
architecture of the TrueNorth \mbox{ASIC-based}
platform~\cite{Merolla668:short} ---, run on an Intel Core i7 CPU
950@3.07~GHz (45~nm CMOS process) with 4 cores and 8 threads, is
5.7~$\mu$J/synaptic event, also in that case excluding the
\mbox{base-line} power consumption.
Table~\ref{tab:uJsynev} summarizes the power consumption of
\dpsnn executed on ARM and Intel against that of the
Compass/TrueNorth simulator.

\begin{table}[!hbt]
\footnotesize
\centering
\caption{Comparison of energetic efficiencies.}
\label{tab:uJsynev}
\begin{tabular}{|c|c|c|}
\hline
\multicolumn{2}{|c|}{\textbf{\dpsnn simulator}} & \multicolumn{1}{|c|}{\textbf{Compass/TrueNorth sim.}} \\
\hline
            \textbf{ARM}           &           \textbf{Intel}                &           \textbf{Intel} \\
\hline
1.1~($\mu$J / syn event)           & 3.4~($\mu$J / syn event)                & 5.7~($\mu$J / syn event) \\
\hline
\end{tabular}
\end{table}

The ARM architecture on Jetson requires about $3\times$ less energy
than Intel, but is about $5 \times$ slower (see ARM \mbox{4-core} row
in Table~\ref{tab:ARM_E2S} 1110~J and 185~s vs 3440~J and 37.4~s, 4
Intel cores in Table~\ref{tab:x86_E2S}).
Moreover, the way lower baseline for ARM makes it an interesting
candidate for clusters that can be populated much more densely than
what is actually possible with Intel.

The profiling of the computation and communication components reported
in Figure~\ref{fig:IntelProfile}, Figure~\ref{fig:TrenzProfile} and
Figure~\ref{fig:JetsonProfile} demonstrated the critical impact of
interconnect on the scaling, limiting the size of the network that can
be simulated in \mbox{real-time}.
The last two rows of Table~\ref{tab:x86_E2S} prove the burden of
interconnect design on the \mbox{energy-to-solution}.
Packets carrying spikes at each simulation step are small, as
quantified in Section~\ref{sec:scaling_real_time}.
The observed effect on scaling is therefore \mbox{latency-related},
not due to lack of bandwidth.

In conclusion, the design of \mbox{low-latency},
\mbox{energy-efficient} interconnects supporting collective
communications is of primary importance to enable a time- and
\mbox{energy-efficient} exchange of neural spikes; this is expected to
not only make cortical simulations possible at a larger scale but also
push their use in embedded systems where it is often precluded by
tight \mbox{real-time} constraints and limited power budget.

\section*{Availability of code and data}
The source code of the \dpsnn engine and the data that support the findings 
of this study are openly available in GitHub at 
https://github.com/APE-group/201812RealTimeCortSim.
The \dpsnn code also corresponds to the internal svn release 1163 of the 
APE group repository.

\section*{Acknowledgment}

This work has received funding from the European Union’s Horizon 
2020 Framework Programme for Research and Innovation under \hbpgrant, 
\exanestgrant and \euroexagrant.




\end{document}